\documentclass[aps,prb,twocolumn,showpacs]{revtex4-1}
\pdfoutput=1
\usepackage{epsfig}
\usepackage{amsmath}
\usepackage{amssymb}

\begin{document}

\title{Non-Fermi Liquid Quantum Impurity Physics from  non-Abelian Quantum Hall States}
\author{Stuart A. Sevier}
\email[sas2933@ucla.edu ]{Current address: Department of Physics and Astronomy, University of California, Los Angeles, California 90095-1547, USA.}
\affiliation{Department of Physics, The University of Texas at Austin, Austin, Texas 78712, USA}
\author{Gregory A. Fiete}
\affiliation{Department of Physics, The University of Texas at Austin, Austin, Texas 78712, USA}

\date{\today}

\begin{abstract}

We study the physics of electron tunneling between multiple quantum dots and the edge of a quantum Hall state.   Our results generalize earlier work [G. A. Fiete, W. Bishara, C. Nayak, Phys. Rev. Lett. {\bf 101}, 176801 (2008)] in which it was shown that a {\em single} quantum dot tunnel coupled to a non-Abelian quantum Hall state can realize a {\em stable} multi-channel Kondo fixed point at low-energy.  In this work, we investigate the physics of multiple dots and find that a rich set of possible low-energy fixed points arises, including those with non-Fermi liquid properties.  Previously unidentified fixed points may also be among the possibilities.  We examine both the situation where the dots are spatially separated and where they are in close proximity.  We discuss the relation to previous work on two-impurity Kondo models in Fermi liquids and highlight new research directions in multiple quantum impurity problems.

\end{abstract}



\maketitle

\section{Introduction}
A quantum impurity is a spatially localized impurity with an internal degree of freedom taking discrete values.  In many cases, the internal degree of freedom is spin, but it could also be an ``orbital" degree of freedom, or even an ``occupation" degree of freedom (as it will be in this work).  Perhaps the most well-known quantum impurity problem is a local moment (spin) inside a metal.  When the effective exchange coupling between the magnetic impurity and the itinerant electrons of the metal is antiferromagnetic, the Kondo effect occurs at low temperatures.\cite{Hewson}  The study of the Kondo effect was originally motivated by an apparent low-temperature divergence of the electrical resistance of metals with dilute magnetic impurities.\cite{Kondo:ptp64}  

In the simplest version of the Kondo problem, a spin-1/2 moment is antiferromagnetically coupled to the local s-wave scattering channel of the impurity.  While at high energies the coupling can often be regarded as perturbative, the low-energy behavior flows to strong coupling where the moment is effectively ``screened out" by a local singlet formation with the spins of the conduction electrons.\cite{Anderson:jpc70}   It has been shown that at low temperatures a type of ``local" Fermi liquid physics arises\cite{Nozieres:jltp74}  where the impurity is characterized by single-particle scattering phase shifts for each spin and these each reach the unitary limit of maximal scattering, $\pi/2$. (The approach to the unitary limit is the cause of the rise in the resistivity at low temperatures in metals with dilute magnetic impurities.)  The local Fermi liquid picture has proved to be of great utility, particularly in the interpretation of an interesting class of scanning tunneling microscope experiments on Kondo impurities.\cite{Fiete:rmp03}

The non-trivial many-body physics of the spin-1/2 Kondo effect\cite{Wilson} spawned an industry of higher spin and multi-channel variants of the Kondo problem.\cite{Hewson,Kondo,Nozieres:jpp80,Emery:prb92,Sengupta:prb94,Ludwig:prl91,Affleck:npb91,Affleck:prb92,Affleck93,Fabrizio:prl95,Andrei:prl95,Andrei:prl84,Tsvelik:zpb84,Borda09,Fiete:prb02,Pustilnik:prl01}
In general, the low-energy fixed points of the multi-channel Kondo models are highly non-trivial and depend on the size of the spin, whether the coupling to all the electron channels is the same or different, and whether the exchange coupling is the same or different for the $x,y$ and $z$-components of the impurity spin.  One of the key results to emerge is that for arbitrary spin, $S$, and isotropic channel and exchange coupling, the low-energy fixed point is {\em non-Fermi liquid} when the number of conduction electron channels, $k$, is greater than $2S$.\cite{Nozieres:jpp80,Affleck:npb90,Affleck:npb91,Affleck_2:npb91} Thus, for $S=1/2$, the $k=2$ channel Kondo model has non-Fermi liquid properties.\cite{Emery:prb92,Sengupta:prb94} The non-Fermi liquid properties are manifest in the impurity susceptibility (and also other quantities) having an anomalous temperature dependence, $\chi_{\rm imp} \propto T^{-(k-2)/(k-2)}$ for $k>2$ and  $\chi_{\rm imp} \propto \ln(T_K/T)$ for $k=2$, where $T_K$ is the Kondo temperature.  The non-Fermi liquid properties only appear when the spin is ``overscreened", {\it i.e.} when there are more conduction electron channels than units of spin-1/2.\cite{Nozieres:jpp80}  (However, it should be emphasized that having $k> 2S$ {\em does not guarantee} non-Fermi liquid physics--that generally requires fine tuning of couplings.\cite{Affleck:prb92, Fabrizio:prl95,Andrei:prl95,Jerez:prb98})  On the other hand, the many-body physics of the ``underscreened" case of $k<2S$ is rather similar to $k=2S$.\cite{Nozieres:jpp80}  Single channel Kondo models have also been studied in Luttinger liquids with 2-channel-like properties realized under some conditions.\cite{Furusaki:prl94,Schiller:prb95,Frojdh:prl95,Frojdh:prb96, Wang:prl96,Egger:prb98,Durganandini:prb06}

A closely related set of studies in the mesoscopic/quantum dot context have shown that the {\em charge fluctuations} on a quantum dot coupled to Fermi liquid leads mimic the spin fluctuations in the Kondo effect, and under some circumstances the combined ``occupation" and spin degrees of freedom can lead to 2-channel, non-Fermi liquid Kondo physics.\cite{Matveev:jetp90,Matveev:prb95,Kim:jpcm03,Sacramento:prb91}  
On the other hand, in certain set-ups involving two quantum dots with charge fluctuations and active spin degrees of freedom, a novel $SU(4)$ Fermi liquid fixed point may be obtained,\cite{Borda:prl03,LeHur:prb03} or even 2-channel behavior.\cite{Oreg:prl03,Zarand:prl06}  Thus, even with Fermi liquid leads, a variety of interesting quantum impurity behaviors occur as a result of charge fluctuations.

An important extension of these works considers a quantum dot coupled to a reservoir (or lead) with interacting electrons.\cite{Sade:prb05,Goldstein08,Goldstein09,Goldstein:prb10,Wachter:prb07,Lerner:prl08}  For example, the interactions in a Luttinger liquid\cite{Giamarchi,Gogolin} lead have been shown to influence the occupation of the dot, with analogies to anti-ferromagnetic and ferromagnetic versions of the Kondo model.\cite{Furusaki:prl02}  However, the {\em charge} fluctuations of a dot are insensitive to the strength of interactions in a Luttinger liquid\cite{Goldstein08} (so long as the effective Kondo model remains antiferromagnetic) and thus are the same as in the Fermi liquid case (generically described by a $k=1$, $S=1/2$ Kondo model when the spin is ignored--due to a polarized state created by the application of a strong magnetic field, for example--and only one reservoir couples to the dot).  Experimentally, the charge fluctuations can be measured capacitively, providing a test for these relations.\cite{Ashoori:prl92,Berman:prl99,Duncan:apl99} 
 
Besides a Luttinger liquid, there are not many gapless non-Fermi liquid phases of interacting itinerant electrons.  However, the edge states of non-Abelian fractional quantum Hall systems are both interacting and fundamentally different from a Luttinger liquid.\cite{Nayak:rmp08} (Another recent example is the ``helical liquid" found at the edge of quantum spin Hall systems,\cite{Hasan:rmp10,Wu:prl06,Xu:prb06,Tanaka:prl09,Zyuzin:prb10}  and some theoretical studies of the Kondo effect in such systems have been reported.\cite{Wu:prl06,Maciejko:prl10,Law:prb10}) Recently it was discovered\cite{Fiete:prl08,Fiete:prb10} that if a quantum dot is tunnel coupled (electron tunneling only) to a non-Abelian quantum Hall state of the Read-Rezayi type,\cite{Read:prb99} then a {\em stable, multi-channel Kondo model} with non-Fermi liquid impurity properties is realized in the charge fluctuations.  Moreover, the usually fragile non-Fermi liquid Kondo fixed point is ``topologically protected"\cite{Fiete:prl08,Fiete:prb10} by the topological order\cite{Nayak:rmp08} of the Read-Rezayi state.  This situation thus provides a rare opportunity to potentially observe in experiment the non-Fermi liquid physics of multi-channel Kondo models with $k>2$, as there are signs of plateaus at filling fractions for $k>2$ that would have Read-Rezayi states as leading candidates.\cite{Nayak:rmp08}  We remark that the argument can also be turned around to show that {\em if} the multi-channel non-Fermi liquid physics is observed, then the quantum Hall state must be a non-Abelian state.\cite{Fiete:prl08,Fiete:prb10}  (A non-Abelian quantum Hall state of any type has not yet been unambiguously observed in experiment.  The observation of a non-Abelian state of any type in any system would be a first in physics.) 

In this work, we extend the single-dot results of Refs.[\onlinecite{Fiete:prl08},\onlinecite{Fiete:prb10}] to include the case of multiple quantum dots. We will focus most of our attention on the double-dot system, which shares many features of a 2-impurity $k$-channel $S=1/2$ Kondo system in a non-Fermi liquid system of itinerant electrons.  To the best of our knowledge, the 2-impurity model has only been studied in a non-interacting electron gas and is known to possess a non-Fermi liquid 2-channel Kondo critical point for a special value of parameters when particle-hole symmetry is present.\cite{Gan:prl95,Gan:prb95,Jones:prl87,Jones:prl88,Affleck:prl92,Affleck:prb95,Ingersent:prl92,Georges:prl95}  In this paper we ask two main questions:  (1) Can the {\em topologically protected} multi-channel nature of each individual impurity\cite{Fiete:prl08,Fiete:prb10} affect the possible phases of the 2-impurity (and more generally, multiple impurity) case?  (2) What is the effect on the phase diagram of the non-Fermi liquid, non-Luttinger liquid itinerant electron gas that mediates the impurity-impurity interaction?  As we will see, these are very difficult questions to answer in general.  However, we are able to obtain some answers in particular limits.  In many cases, the multiple dot case can be shown to reduce to the single dot case.  In other cases, there is a strong possibility for previously undiscovered fixed points with non-Fermi liquid properties.  We hope our work will provide a strong motivation for opening up a new line of inquiry in quantum impurity problems (in addition to shedding more light on the non-Abelian quantum Hall states themselves).

Our paper is thus organized as follows.  In Sec.~\ref{sec:QH} we cover some necessary essentials of the theory of quantum Hall edges that we will use in our discussion.  In Sec.~\ref{sec:multi} we discuss the coupling of multiple dots in several geometries and limits.  Particular attention is paid to the case of two dots.  Finally, in Sec.~\ref{sec:conclusions} we discuss our conclusions and suggest possible extensions of the present work.

\section{Quantum Hall Edge Physics}
\label{sec:QH}

In this section, we provide a brief overview of the relevant aspects of quantum Hall edge physics needed for our study.  This section should be helpful to those with a background in quantum impurity models, but who are less familiar with quantum Hall systems.  We have tried to include enough detail to keep the paper self-contained.  However, should more details be desired, a more extended discussion with the same notional conventions can be found in Ref.[\onlinecite{Fiete:prb10}].

\subsection{Fundamentals and The Current State of Experiments}
Quantum Hall systems fall into two broad categories:  (i) Abelian and (ii) non-Abelian.\cite{Nayak:rmp08}  Quasi-particle and quasi-hole excitations in quantum Hall systems are generally gapped in the bulk; the naming of Abelian and non-Abelian refers to the braiding properties of these excitations above the gap.\cite{Wen}  If the braiding operations of particles commute, the state is Abelian, while if they do not commute the state is non-Abelian\cite{Stern:nat10} (the latter generally requires a non-trivial ground state degeneracy beyond that of the Abelian case).   All the integer quantum Hall states\cite{Halperin:prb82} are Abelian, as well as the celebrated Laughlin states.\cite{Laughlin83}  (The Laughlin states possess anyon excitations with exchange angles different from both fermions and bosons.\cite{Arovas:prl84})  The huge families of hierarchy states are Abelian, too.\cite{Haldane83,Halperin84,Jain89,Wen95}  Thus, non-Abelian states appear to be rather rare.  However, they can readily be constructed from correlation functions\cite{Read:prb99,Moore:npb91,Greiter92,Nayak96c,Bonderson08,Bernevig:prl09,Hansson:prl09,Hansson:prb07,Read07} in a certain class of conformal field theories (CFT).\cite{CFT}  

The construction of these wavefunctions would be nothing more than an excercise in mathematical physics if there were not strong numerical\cite{Storni:prl10,Arkadiusz:prl10,Morf98,Rezayi00,Wang09,Bonderson09,Feiguin:prb09,Wojs:prb09,Rezayi:prb09,Wang09} as well as experimental evidence\cite{Xia:prl10,Chickering:prb10,Dolev:nat08,Miller:nap07,Radu08,Willett:pnas09,Dolev:prb10} that non-Abelian states are preferred by nature under certain conditions. In the GaAs/AlGaAs samples most commonly used, these conditions are best met in the second Landau level\cite{Rezayi:prb09,Wang09} where non-Abelian candidates tend to have the largest gaps.\cite{Xia04,Choi07,Dean:prl08}  However, it remains an open challenge to provide a compelling demonstration of a non-Abelian state experimentally.  Fortunately, the number of proposals for experiments capable of detecting non-Abelian properties has grown dramatically in recent years.\cite{Nayak:rmp08,Fiete:prl08,Fiete:prb10,Bonderson:prb10,Gervais:prl10,Yang:prb09,Wan:prb08,Hu:prb09,Seidel:prb09,Chen:09,Law:prb08,Wang:prb10,Overbosch,Sarma:prl05,Rosenow:prl08,Bishara09,Bishara_2_09,Rosenow09,Stern:prl06,Bonderson:prl06,Bonderson_2:prl06,Cooper:prl09,Grosfeld:prb06,Ilan:prl08,Grosfeld:prl06,Grosfeld:prl09,Bishara:prb08,Law:prb08,Bena:prb06,Fendley09,Fendley:prl06,Fendley:prb07}

\subsection{Key Elements of Quantum Hall Edge Theories}

In this work, we will assume that a certain class of non-Abelian states exist in nature, the Read-Rezayi (RR) states,\cite{Read:prb99} and explore their consequences for electron tunneling to multiple quantum dots.  The RR states are all {\em spin polarized} and relevant to filling fractions $\nu=k/(k+2)$ (or $k=2+k/(k+2)$ when the lowest Landau levels are inert).   The simplest of the non-Abelian states is the $k=2$ Moore-Read (MR) Pfaffian\cite{Moore:npb91} and its particle-hole conjugate,\cite{Levin07,Lee07,Bishara:prbR08} which are leading candidates for filling fraction $\nu=5/2$. The filling fraction $\nu=12/5$ state may be in the same class as the particle-hole conjugate of the $k=3$ Read-Rezayi state\cite{Bishara:prbR08} (or possibly the in Bonderson-Slingerland hierarchy\cite{Bonderson08,Bonderson09}).  If a clear signature of a state eventually appears at $\nu=13/5$, then the $k=3$ RR state would be a leading candidate. We note that the Laughlin $\nu=1/3$ state is the only Abelian RR state, realized for $k=1$.  We remind the reader that while the bulk of quantum Hall states are generally gapped to single-particle excitations, the edges are gapless and form a quasi-one dimensional system.  Due to the breaking of time-reversal symmetry in the quantum Hall states, the edges have a definite directionality associated with them and are often referred to as ``chiral".  It should be noted, however, that while there is a net direction to current and heat flow along the edge (not always in the same direction\cite{Kane94,Kane:prb95,Levin07,Lee07,Bishara:prbR08}), there is also the possibility of counter-propagating edge currents that only partially cancel each other.\cite{Chklovskii:prb92,MacDonald:prl90,Wen:prb90,Wen92,Wen:ijmp92,Wen95}

As was emphasized in Refs.[\onlinecite{Fiete:prl08},\onlinecite{Fiete:prb10}], the key element that unifies the multi-channel Kondo models, candidate non-Abelian RR wavefunctions, and the gapless edge theory of non-Abelian RR states is the underlying $\mathbb{Z}_k$ parafermion CFT.  The parafermion theory is obtained from the coset construction\cite{CFT}  $SU(2)_k/U(1)$, and its Lagrangian is given by an $SU(2)_k$ chiral WZW model in which the $U(1)$ subgroup has been gauged.\cite{Karabali90}  [See Eq.\eqref{eq:S-RR}.]  For our applications, we will mainly need the expressions for edge electron operator in different quantum Hall states, and the scaling dimensions of the fields that enter it (which are set by the action of the corresponding edge theory).  Working from the simplest edges in the RR series, to the general case we have the following expressions.

The low-energy edge theory of the ($k=1$ RR or) $\nu=1/3$ Laughlin state (and for general $\nu$, including $\nu=1$) can be expressed in terms of bosonic fields and has an action that takes the form
\begin{equation}
\label{eq:S_L}
S_{\rm edge}^{\rm Laugh}=\frac{1}{4\pi \nu}\int d\tau dx\, \partial_x \phi (i\partial_\tau   + v \partial_x) \phi ,    
\end{equation}
where the chiral bosonic fields satisfy the commutation relations $[\phi(x'),\phi(x)]=i \pi \nu {\rm sgn}(x-x')$; and $v$ is the velocity of the edge mode, determined by non-universal properties of the edge confining potential and interactions. (Throughout our paper we have set $\hbar=1$.) 

The electron operator on the edge is 
\begin{equation}
\label{eq:Psi_Laugh}
\Psi_{e,{\rm Laugh}}^\dagger=e^{i\phi/\nu},
\end{equation}
and the quasi-particle operator of charge $\nu e$ is $\Psi_{q,\rm Laugh}^\dagger=e^{i\phi}$ (which we do not use in this paper).

The MR state, like the Laughlin states, is spin polarized.
Its edge action is the sum of the neutral and charged sectors,
$S=S_n+S_c$, with
\begin{equation}
\label{eq:S_n}
S_n^{\rm MR}=\int d\tau dx \; i\psi \left(i\partial_\tau \psi + v_n \partial_x \psi\right),    
\end{equation}
and
\begin{equation}
\label{eq:S_c}
S_c=\frac{2}{4\pi}\int d\tau dx \partial_x \phi\left(i\partial_\tau \phi  + v_c \partial_x \phi\right) ,    
\end{equation}
where $v_n$ is the neutral mode velocity and $v_c$ is the charge mode velocity. Typically $v_n<v_c$, because of the repulsive interactions in the charge sector.\cite{Wan:prl06,Wan:prb08}   Note that \eqref{eq:S_c} is just \eqref{eq:S_L} with $\nu=1/2$ and $v=v_c$. Here $\psi$ is a Majorana fermion, the $\mathbb{Z}_2$ parafermion.

The electron creation operator in the MR state is:
\begin{equation}
\label{eq:Psi_MR}
\Psi_{e,{\rm MR}}^\dagger=\psi e^{i2\phi},
\end{equation}
where we have used the convention ${\rm dim}[e^{i\alpha \phi}]=\nu \frac{\alpha^2}{2}$ [follows from \eqref{eq:S_L}] and ${\rm dim}[\psi]=1/2$, so that ${\rm dim}[\Psi_{e,{\rm MR}}]=3/2$.  Note that the electron operator \eqref{eq:Psi_MR} is a combination of a Majorana fermion and an exponential of the boson, identical to that appearing in \eqref{eq:Psi_Laugh}.    

The general RR states are also spin polarized, and their edge action is again the sum of neutral and charged sectors. The charge sector of the RR edge theory (neglecting the 2 filled lower Landau levels if one is considering a state in the second Landau level) is identical to \eqref{eq:S_L} with $\nu=k/(k+2)$.  The neutral $\mathbb{Z}_k$ parafermion sector is given by\cite{Karabali90}
\begin{multline}
\label{eq:S-RR}
S_{{SU(2)_k}/U(1)} =  \frac{k}{16\pi}\int d\tau dx\,
\text{tr}\left( {\partial_x} {g^{-1}}
\overline{\partial} g\right)
\\ -\:
i\frac{k}{24\pi}\int dx d\tau dr\,  \epsilon^{\mu\nu\lambda}
\text{tr}\left( {\partial_\mu} g\,{g^{-1}}
{\partial_\nu} g\,{g^{-1}}\,{\partial_\lambda}g\,{g^{-1}}\right)
\\
 + \frac{k}{4\pi} \int dx d\tau \, \text{tr}
\Bigl({A_x}\overline{\partial}g \cdot g^{-1}
- \overline{A}g^{-1}{\partial_x}g
+ {A_x}g  \overline{A}g^{-1}
- {A_x} \overline{A} 
\Bigr),
\end{multline}
where $\overline{\partial}\equiv i{\partial_\tau} + {v_c}{\partial_x}$, $\overline{A}\equiv A_{\tau}-i{v_n}{A_x}$, and the field $g$ takes values in $SU(2)$.
The second integral in Eq.\eqref{eq:S-RR} is over any three-dimensional manifold $M$
which is bounded by the two-dimensional spacetime of the edge $\partial M$.
The value of this integral depends only on the values of the field $g$
at the boundary $\partial M$. The gauge field has no Maxwell term,
so its effect is to set to zero the $U(1)$ current to which it is coupled.

The electron operator in the RR state is obtained by combining
appropriate fields from the charge and neutral sectors:\cite{Read:prb99}
\begin{equation}
\label{eq:Psi_RR}
\Psi_{e,{\rm RR}}^\dagger=\psi_1 e^{i\frac{k+2}{k}\phi},
\end{equation}
where $\psi_1$ is the $\mathbb{Z}_k$ parafermion field with ${\rm dim}[\psi_1]=1-1/k$.  Thus, using the conventions in \eqref{eq:S_L} with $\nu=k/(k+2)$, we have ${\rm dim}[\Psi_{e,{\rm RR}}]=3/2$, independent of $k$.

\subsection{Connection to Multichannel Kondo Models}
 
Conformal field theory methods offer an elegant solution to the $k$-channel Kondo models in Fermi liquids.\cite{Affleck:npb90,Affleck:npb91,Affleck_2:npb91}  The conduction electron Hamiltonian is expressed as a sum of three terms (referred to as a ``conformal embedding'') that describe ``charge", ``spin", and ``flavor" sectors.  These sectors possess current operators with $U(1)$, $SU(2)$, and $SU(k)$ symmetry, respectively.  Specifically, for the $k$-channels of free electrons $\psi_i$ described by $H_0$, one has
\begin{equation}
\label{eq:H_conf_decomp}
\sum_{i=1}^kH_0[\psi_i]=H[U(1)]+H[SU(2)_k]+H[SU(k)_2].
\end{equation}
Conformal embedding is useful for the $k$-channel Kondo model because it is only the $SU(2)_k$ currents that couple to the local moment $\vec S$: the conduction electron terms with $U(1)$ and $SU(k)$ symmetry thus play no role in the quantum impurity physics.\cite{Affleck:npb90,Affleck:npb91,Affleck_2:npb91}  These $SU(2)_k$ currents can be realized at the operator level in terms of a chiral boson $\varphi$ and a parafermion field $\psi_1$ [the same one in \eqref{eq:Psi_RR}]:
\begin{eqnarray}
\label{ParafCurrents}
J^+ &=& \sqrt{k}\psi_1 e^{i\varphi},\cr
J^{-} &=& \sqrt{k}\psi_1^{\dagger} e^{-i\varphi},\cr
J^z &=& \frac{k}{2}\partial_x \varphi,
\end{eqnarray}
where the chiral boson is normalized so that ${\rm dim}[e^{i\varphi}]=1/k$.  The important relation to note is that the electron creation operator \eqref{eq:Psi_RR} and $J^+$ in \eqref{ParafCurrents} only differ in the normalization of the bosonic field. In other words, the edge electron operator on a RR state and the $SU(2)_k$ currents are closely related objects.  

In the problem of electron tunneling to a {\em single} quantum dot, the electron operator can be ``rotated" into an $SU(2)_k$ current operator with a unitary transformation, providing an {\em exact} mapping to a $k$-channel Kondo model.\cite{Fiete:prl08,Fiete:prb10} Thus, the {\em stable} multi-channel Kondo models depend crucially on a non-Abelian quantum Hall edge state having a {\em unique} electron operator, which does {\em not} occur for the particle-hole conjugate RR states.\cite{Fiete:prl08,Fiete:prb10} As the particle-hole conjugate RR states generally result in Fermi liquid impurity physics, we do not consider them here.\cite{Fiete:prl08,Fiete:prb10}
 
\section{Multiple Quantum Dots coupled to Quantum Hall States}
\label{sec:multi} 
 
\begin{figure}[b]
\includegraphics[width=0.85\linewidth,clip=]{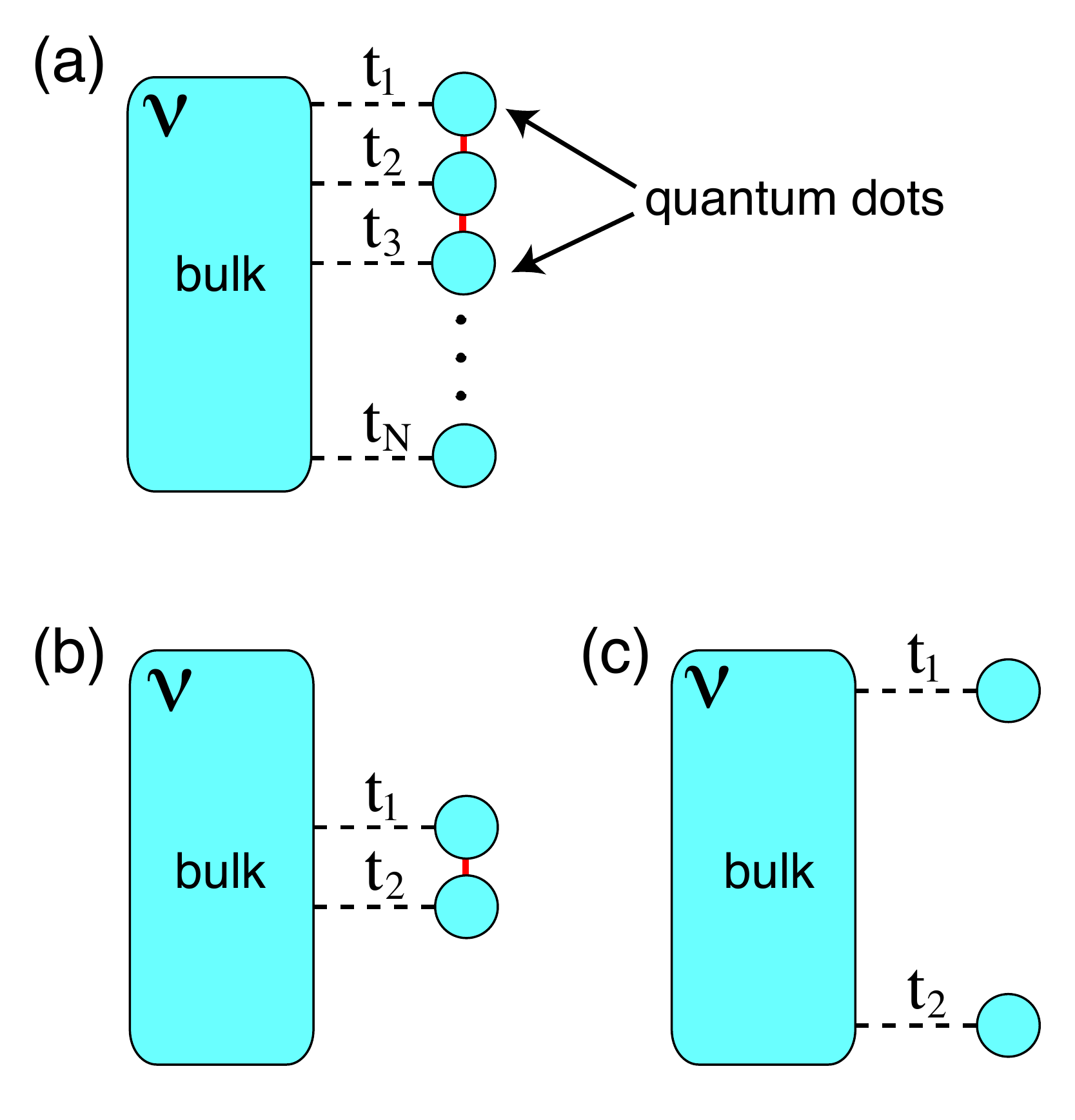}
 \caption{\label{fig:schematic} (color online) Schematic of multiple quantum dots edge (tunnel) coupled to a quantum Hall system as described in the text. (a) Illustrates the general situation of an arbitrary number of dots, while (b,c) illustrate the different cases of two quantum dots.  In (b) the dots are close enough together to directly interact with each other, while  in (c) the dots are far enough apart that there is no direct interaction between them. Side gates (not shown) would enable one to adjust the spectrum on individual dots to a degeneracy point, $E(N)=E(N+1)$, where number fluctuations are enhanced.  The quantum Hall state is assumed to be at filling fraction $\nu=2+k/(k+2)$ and described by a state in the Read-Rezayi series.  As in earlier work\cite{Fiete:prl08,Fiete:prb10} we assume the lower Landau levels are inert and focus our study on the partially filled Landau level at filling fraction $\nu=k/(k+2)$. The electron tunneling amplitudes between dot $i$ and the edge are denoted by $t_i$.}
 \end{figure}

 An important motivation to study quantum dots in the quantum Hall context comes from the realization that finite-size effects are influenced by the underlying bulk quantum Hall state.\cite{Fiete:prl07,Fiete:prl08,Fiete:prb10,Sarma:prl05,Stern:prl06,Bonderson:prl06,Bonderson_2:prl06,Feldman:prl06,Feldman:prb07,Bonderson:prb10,Wan:prl06,Hu:prb09,Chamon:prb97,Rosenow:prl08,Bishara09,Bishara_2_09,Rosenow09,Grosfeld:prb06,Ilan:prl08,Law:prb08}  In this work we focus our attention on the {\em charge} fluctuations of multiple quantum dots arranged along the edge of a quantum Hall system as shown in Fig.\ref{fig:schematic}.  The charge fluctuations on the quantum dots can be measured capacitively.\cite{Ashoori:prl92,Berman:prl99,Duncan:apl99}  Because the quantum Hall states we consider are spin polarized, the spin degree of freedom is assumed to be absent in all our considerations.

\subsection{Multi-dot Hamiltonian}

The Hamiltonian for our system is given by 
\begin{equation}
\label{eq:H}
H=H_{\rm edge}+H_{\rm QD}+H_{\rm tun}+H_{\rm int},
\end{equation}
where $H_{\rm edge}$ describes the gapless excitations of the quantum Hall edge, generally taken to be a RR state at filling $\nu=k/(k+2)$ (with any completely filled Landau levels assumed inert\cite{Fiete:prl08,Fiete:prb10}), $H_{\rm QD}$ describes the states on the quantum dots, including any direct interactions or couplings between them, $H_{\rm tun}$ describes the tunneling of electrons between the quantum Hall edge and the quantum dots, and $H_{\rm int}$ describes the interactions between the electrons on the quantum dots and the electrons on the quantum Hall edge.  Note that quantum dots are assumed spatially well separated from the quantum Hall states so that quasi-particle tunneling is not allowed, as it would be in point contact geometry.\cite{Chamon:prb97,Fendley:prl06}
 
As we are interested in the physics of charge fluctuations on the quantum dots due to interactions with the quantum Hall edge and other quantum dots, it is important to assume that each quantum dot (if it were isolated) is close to a number degeneracy point: $E(N)=E(N+1)$, where $N$ is the number of electrons on the dot.  Here $E(N)$ is the energy of a dot with $N$ electrons on it.  This means that electron number fluctuations are energetically allowed in the presence of weak electron tunneling.  We assume the dots are small enough that we can truncate their Hilbert space to just a single level at the Fermi energy with the same energy for single or no occupancy.  The other levels are assumed to be further away in energy than the temperature or any energy scale associated with tunneling.

Under these conditions, we consider the following form of the quantum dot Hamiltonian,
\begin{equation}
\label{eq:H_QD}
H_{\rm QD}=\sum_i H_{{\rm QD},i}+\sum_{i<j}(t_{ij}d^\dagger_id_j +{\rm H.c})+\sum_{i<j}V_{ij}n_in_j,
\end{equation}
where $H_{{\rm QD},i}=\epsilon_i d^\dagger_id_i$ and  $d_i$ ($d_i^{\dagger}$) is the fermionic annihilation (creation) operator of electrons on dot $i$. The number operator $n_i=d^\dagger_id_i$.  At the degeneracy point of each isolated dot, $E(N)=E(N+1)$, which implies $\epsilon_i=0$.  The amplitude $t_{ij}$ describes direct electron tunneling between dots $i$ and $j$, and $V_{ij}$ a direct density-density interaction between the same dots. 

The Hamiltonian describing electron tunneling between the dots and the edge is
\begin{equation}
\label{eq:H_tun}
H_{\rm tun}=\sum_i t_i \Psi^\dagger_e(x_i)d_i+{\rm H.c.},
\end{equation}
where $t_i$ is the tunneling amplitude to dot $i$ at position $x_i$ along the edge, and $\Psi^\dagger_e(x_i)$ is the edge electron creation operator at position $x_i$.  Depending on the quantum Hall state under consideration, the electron creation operator is given by \eqref{eq:Psi_Laugh}, \eqref{eq:Psi_MR}, or \eqref{eq:Psi_RR}.

The interactions between the quantum Hall edge charge and the quantum dots is described by
\begin{equation}
\label{eq:H_int}
H_{\rm int}=\sum_i V_i d^\dagger_id_i \partial_x \phi(x_i),
\end{equation}
where the local density of electrons on the quantum Hall edge at position $x_i$, $\Psi_e^\dagger(x_i)\Psi_e(x_i)$, is proportional to $\partial_x \phi(x_i)$ of the charge sector alone,\cite{Giamarchi} and $V_i$ is the strength of the interaction.  Note that our notational conventions are double subscripts for dot-dot terms ($t_{ij},V_{ij}$) and single subscripts for edge-dot terms ($t_i,V_i$).

For the assumptions under which we are working, the expressions (\ref{eq:H})-(\ref{eq:H_int}) along with the possible edge theories \eqref{eq:S_L}, \eqref{eq:S_n}, and \eqref{eq:S-RR} provide a rather general description.  We are now ready to study the Hamiltonian in a number of different limits, focussing on the charge fluctuations in the case of two quantum dots.

\subsection{Double-dot Hamiltonian}

A great deal of the structure of the fixed points of the multi-dot case is revealed by the double-dot case.  We will focus on the two limits shown in Fig.\ref{fig:schematic}(b,c).  Let us first consider what type of behaviors  we should expect based on physical considerations alone.  In Fig.\ref{fig:schematic}(b) the dots are close enough to interact with each other directly, {\it i.e.} $t_{ij},V_{ij}\neq 0$.  So, if the dots are more strongly coupled to each other than the edge, we might expect the dots to behave as a ``composite" dot with a higher effective ``spin" than the single-dot case.\cite{Fiete:prl08,Fiete:prb10}  On the other hand, if the dots are far enough apart not to interact directly,  {\it i.e.} $t_{ij}=V_{ij}=0$ as shown in Fig.\ref{fig:schematic}(c), then their only interactions will be those mediated by the quantum Hall edge states.  In this picture, one expects some variant of RKKY-type interactions,\cite{Ruderman:rmp54} at least at higher energy scales (although still small compared to the quantum Hall bulk gap, for example).  Even for well-separated dots, in the low-energy limit they effectively ``collapse" to a single point (because the characteristic wavelength of edge excitations will eventually be much larger than the inter-dot spatial separation) and the physics will be similar (in some limits) to that of two nearby but non-interacting quantum dots.  

Thus, there are two primary limits to consider: (i) The composite ``dot" and (ii) the situation where RKKY-type indirect mediated interactions are important.\cite{Ruderman:rmp54}  In each case, we would like to understand the behavior of the charge fluctuations on the dots near their degeneracy points, and we would like to draw parallels to the previously studied 2-impurity Kondo problem in a Fermi liquid.  In particular, we would like to know if there are any unusual quantum impurity features or behaviors in this scenario.

\subsubsection{Composite dot}

In the composite dot limit, the quantum dots can be taken to be at the same spatial location, $x_1=x_2=0$, from the point-of-view of the edge electrons.  The quantum dot, tunneling, and interaction Hamiltonians then take the form
\begin{equation}
\label{eq:H_QD_2}
H_{\rm QD}=\epsilon_1 n_1 +\epsilon_2 n_2+t_{12}d^\dagger_1d_2 +t_{12}^*d^\dagger_2d_1+V_{12}n_1n_2,
\end{equation}
\begin{equation}
H_{\rm tun}=\Psi_e^\dagger(0)(t_1d_1+t_2d_2)+{\rm H.c.},
\end{equation}
and
\begin{equation}
H_{\rm int}=(V_1 n_1+V_2 n_2)\partial_x\phi(0).
\end{equation}

If the dots are in close proximity, one would expect $V_{12}$ to be a large energy scale in the problem.  In this case, the dot system would prefer to minimize the energy of the term $V_{12}n_1n_2$ in $H_{\rm QD}$.  This effectively restricts the double-dot Hilbert space to either one or zero total electrons on both dots.  Diagonalizing $H_{\rm QD}$ within the one-electron subspace one finds energy eigenvalues $E_\pm =(\epsilon_1+\epsilon_2)/2\pm \sqrt{(\epsilon_1-\epsilon_2)^2+4|t_{12}|^2}/2$, and the Hamiltonian can be written as
\begin{equation}
\label{eq:QD_trans}
H_{\rm QD}^{\rm 1-elec}=E_- d_-^\dagger d_- +E_+ d_+^\dagger d_+,
\end{equation}
where $ d_-,d_+$ are the diagonal basis of \eqref{eq:H_QD_2}.   The two basis are related via the transformation
\[
\left(
\begin{array}{cc}\label{eq:trans_2}
\frac{t_{12}}{\sqrt{(E_- -\epsilon_1)^2+t_{12}^2}} & \frac{E_+-\epsilon_2}{\sqrt{(E_+ -\epsilon_2)^2+t_{12}^2}}  \\
\frac{E_- -\epsilon_1}{\sqrt{(E_- -\epsilon_1)^2+t_{12}^2}} &  \frac{t_{12}}{\sqrt{(E_+ -\epsilon_2)^2+t_{12}^2}}  \\
 \end{array}
\right)
\Biggl(\begin{array}{c} d_-\\ d_+ \end{array} \Biggr)
=\Biggl(\begin{array}{c} d_1\\ d_2 \end{array} \Biggr),
\]
where we have assumed $t_{12}$ is real without loss of generality. 

If one tunes $\epsilon_1,\epsilon_2$, and $t_{12}$ so that $E_-=0$ then the state of no electron on the level $ d_-$ is degenerate with the state of one electron on the same level.  (One could alternatively take $E_+=0$ if $\epsilon_1+\epsilon_2<0$, for example.) This occupancy degeneracy is a necessary condition for the realization of $S=1/2$ Kondo physics in the double-dot charge fluctuations.  To see if it is sufficient, we must inspect the form of $H_{\rm tun}$ and $H_{\rm int}$ in the diagonal basis to see if the mapping of Refs.[\onlinecite{Fiete:prl08},\onlinecite{Fiete:prb10}]  still goes through. The tunneling Hamiltonian becomes
\begin{equation}
\label{eq:H_tun_2}
H_{\rm tun}=\Psi_e^\dagger(0)(t_- d_- + t_+ d_+)+{\rm H.c.},
\end{equation}
where 
\begin{eqnarray}
t_-&=&\frac{t_1 t_{12}+t_2(E_--\epsilon_1)}{\sqrt{(E_- -\epsilon_1)^2+t_{12}^2}}    
,\\  
t_+&=&\frac{t_1(E_+-\epsilon_2)+t_2t_{12}}{\sqrt{(E_+ -\epsilon_2)^2+t_{12}^2}}
,
\end{eqnarray}
for general values of $E_\pm$.  Note that for generic values of parameters (including $E_-=0$), $t_+\neq0$, so that the level $d_+$ is tunnel coupled to the quantum Hall edge via \eqref{eq:H_tun_2}.  However, if parameters $\epsilon_1,\epsilon_2$, and $t_{12}$ are such that $E_+ \gg E_-=0$, then we may still neglect the level $d_+$ at the same level of approximation that we have neglected levels far away from the Fermi energy in the isolated quantum dot.  At low temperatures this is justified because the occupation of the level $d_+$ will be exponentially close to zero, while the occupation of the level $d_-$ will be $1/2$ on average when $E_-=0$.  Thus, we can safely neglect the $d_+$ term in \eqref{eq:H_tun_2} {\em provided} there is no coupling between $d_-$ and $d_+$ in the transformed $H_{\rm int}$.

Under the transformation just below Eq.~\eqref{eq:QD_trans}, the interaction Hamiltonian becomes
\begin{equation}
\label{eq:H_int_2}
H_{\rm int}=\left(V_- n_-+V_+ n_++V_\pm (d^\dagger_-d_++d^\dagger_+d_-)\right)\partial_x\phi(0),
\end{equation}
where $n_-=d^\dagger_-d_-$, $n_+=d^\dagger_+d_+$, and
\begin{eqnarray}
V_-&=&\frac{V_1 t_{12}^2+V_2(E_--\epsilon_1)^2}{(E_- -\epsilon_1)^2+t_{12}^2}\;,\\
V_+&=&\frac{V_1 (E_+-\epsilon_2)^2+V_2 t_{12}^2}{(E_+ -\epsilon_2)^2+t_{12}^2}\;,\\
V_\pm&=&\frac{V_1t_{12}(E_+-\epsilon_2)+V_2 t_{12}(E_--\epsilon_1)}{\sqrt{(E_- -\epsilon_1)^2+t_{12}^2}\sqrt{(E_+ -\epsilon_2)^2+t_{12}^2}}\;.
\end{eqnarray}
Note that in \eqref{eq:H_int_2} the transformation to $d_-,d_+$ couples the two states and will therefore destroy the exact mapping of the charge fluctuations on the $d_-$ (when $E_-=0$) to the Kondo model {\em unless} $V_\pm=0$.  However, this condition is likely to be approximately satisfied in practice.  We first note that if $V_1=V_2$ (as it would be for identical dots in isolation), then $V_\pm=0$.  However, for similar dots one expects $V_1=V_2+\delta V$ for some small $\delta V$. Doing perturbation theory in the terms $\sim \delta V(d^\dagger_-d_++d^\dagger_+d_-)\partial_x\phi(0)$ will lead to a renormalization of $V_-$ and $V_+$ going as $\delta V^2/(E_+-E_-)n_-n_+(\partial_x\phi(0))^2$.  As the $d_+$ level can safely be ignored under our assumption of $E_+\gg E_-=0$, $V_-$ will obtain a correction going as $\delta V^2/(E_+-E_-)\langle n_+\rangle\langle \partial_x\phi(0)\rangle$, which is clearly rather small as both $\delta V^2/(E_+-E_-)$ and $\langle n_+\rangle$ are independently small.  

Therefore, in the  composite dot limit, when we have tuned $E_+\gg E_-\approx0$, the effective Hamiltonian is
\begin{eqnarray}
H_{\rm comp}&\approx& H_{\rm edge}+E_-n_-\nonumber \\
&+&\left(t_-\Psi^\dagger_e(0)d_-+{\rm H.c}\right)+V_-n_-\partial_x\phi(0),
\end{eqnarray}
where $\Psi^\dagger_e(0)$ is given by one of  \eqref{eq:Psi_Laugh}, \eqref{eq:Psi_MR}, or \eqref{eq:Psi_RR}.  Let us consider the general case of \eqref{eq:Psi_RR}.   Applying the unitary transformation $U=e^{i \alpha S^z \phi(0)}$ with $\alpha=\frac{k+2}{k}-\frac{\sqrt{2(k+2)}}{k}$ gives 
\begin{eqnarray}
\label{eq:H_U_Kondo}
UH_{\rm comp}U^\dagger&=&H_{\rm edge}+hS^z\nonumber \\&+&\lambda_\perp(J^+S^-+J^-S^+) 
+\lambda_zJ^zS^z,\;\;\;
\end{eqnarray}
where 
\begin{equation}
\label{eq:SU2_currents}
J^+=\sqrt{k}\psi_1 e^{i \beta \phi},\;J^-=\sqrt{k}\psi_1^\dagger e^{-i \beta \phi},\;
J^z={k \over 2}\beta \partial_x \phi,
\end{equation}
with $\beta=\sqrt{2(k+2)}/k$, ${\lambda_\perp}=t_-$,
$\beta {\lambda_z}=V_--{v_c}\alpha$, and $h=E_-$. The local spin operators are related to the local $d_-$ composite dot level as: $S^+ =\eta d_-^\dagger, S^-=d_-\eta$,
and $S^z=n_- -1/2$, where $\eta$ are Klein factors that ensure the proper commutation relations are achieved.\cite{Klein_factors}  The $\eta$ have the property that $\eta^\dagger=\eta$, $\eta^2=1$ and they anti-commute with fermions, {\it i.e}. their properties are like non-dynamical Majorana fermions.  For $V_->{v_c}\alpha$, this is the antiferromagnetic Kondo problem, which has an intermediate coupling fixed point.\cite{Fiete:prl08,Fiete:prb10}  

We have thus established that the low-energy physics of a composite double-dot (with $E_+\gg E_-\approx0$ and $V_1-V_2$ small) coupled to a RR non-Abelian quantum Hall state maps onto the problem of a single quantum dot coupled to the same state in a certain range of parameters.  Therefore, the composite dot system will realize a {\em stable} $k$-channel single-impurity non-Fermi liquid Kondo physics near a degeneracy point at filling fraction $\nu=2+k/(k+2)$ (with the completely filled Landau levels assumed inert).\cite{Fiete:prl08,Fiete:prb10}  

If multiple dots are arranged in close proximity, then the preceding analysis can be generalized, and it is clear that one can in general tune to a degeneracy point $E(N)=E(N+1)$ on the multi-dot system as well and realize the physics of a {\em single} dot.  In this sense, a great deal of the Kondo-like  physics of the charge fluctuations on proximate multi-dots is captured by the effective $S=1/2$ single-dot limit studied earlier.\cite{Fiete:prl08,Fiete:prb10}  However, the multi-dot case also allows for the possibility of an effective dot degeneracy greater than 2, which as we will see in the next section corresponds to higher spin Kondo models.

\subsection{Two spatially remote dots}
Here we discuss the situation shown schematically in Fig.\ref{fig:schematic}(c) where at ``high" energies the dots do not directly interact with each other. The only interactions between the dots are those mediated by the gapless edges of the quantum Hall systems.  This will be roughly analogous to the famous RKKY interactions between two remote impurities in a Fermi liquid.\cite{Ruderman:rmp54}

\subsubsection{A candidate low-energy fixed point}
Before we turn to the indirect interactions present at ``high" energies, we can immediately determine one candidate for the low-energy fixed point. We note that regardless of the initial separation of the dots, at sufficiently low energies they reside at the same spatial location from the point-of-view of the edge states.  Thus, the tunneling Hamiltonian \eqref{eq:H_tun} will again effectively have $x_1=x_2=0$ at the fixed point.  However, since the dots are ``far" from each other, one has $t_{ij}=V_{ij}=0$ in \eqref{eq:H_QD} because direct inter-dot tunneling and interactions can be neglected.  This immediately leads to the following effective Hamiltonian of two remote dots at low energies:
\begin{eqnarray}
\label{eq:2_iso}
H=H_{\rm edge}+\left(\Psi_e^\dagger(0)(t_1d_1+t_2d_2)+{\rm H.c.}\right)
\nonumber \\
+\epsilon_1n_1+\epsilon_2n_2 +(V_1 n_1+V_2 n_2)\partial_x\phi(0).\;\;
\end{eqnarray}
Notice that the form of \eqref{eq:2_iso} is exactly of the same form as the combination \eqref{eq:QD_trans}, \eqref{eq:H_tun_2}, and \eqref{eq:H_int_2} with $V_\pm=0$.  (Recall, however, that the Hamiltonians  \eqref{eq:QD_trans}, \eqref{eq:H_tun_2}, and \eqref{eq:H_int_2} were derived for the 0 or 1-particle occupation on the two dots, while \eqref{eq:2_iso} is valid within the 0,1 {\em and} 2-particle subspace on the two isolated dots.)  Thus, much of the previous analysis carries over.  For example, if $\epsilon_1=0$ and $|\epsilon_2| \gg 0$ (or the other way around) in \eqref{eq:2_iso}, then we expect from our previous analysis that the low-energy physics will be that of a single isolated quantum dot: $S=1/2$, $k$-channel non-Fermi liquid Kondo physics.\cite{Fiete:prl08,Fiete:prb10}

An important special case occurs when the two dots and their couplings to the edge are identical: $\epsilon_1=\epsilon_2=\epsilon$, $t_1=t_2=t$, and $V_1=V_2=V$:
\begin{eqnarray}
\label{eq:2_iso_ident}
H=H_{\rm edge}+\left(t\Psi_e^\dagger(0)(d_1+d_2)+{\rm H.c.}\right)
\nonumber \\
+\epsilon(n_1+n_2 )+V( n_1+n_2)\partial_x\phi(0),\;\;
\end{eqnarray}
which, because of the occupation subspace of 0, 1, or 2 particles, maps onto an $S=1$ Kondo model (which has three allowed $S^z$ components).  By using $S^+ =S_1^++S_2^+=\eta_1 d_1^\dagger+\eta_2d_2^\dagger,\; S^-=d_1\eta_1+d_2\eta_2$, and $S^z=S_1^z+S_2^z=n_1+n_2 -1$, where $\eta_1,\eta_2$ are Klein factors that ensure the proper commutation relations are achieved: $[S^+,S^-]=2S^z$,\cite{Klein_factors} the Hamiltonian is again cast into the form \eqref{eq:H_U_Kondo} with $S=1$.  Note that in the general situation $\lambda_\perp \neq \lambda_z$, so that the Hamiltonian is {\em exchange} anisotropic.  (It remains {\em channel} isotropic by virtue of the unique edge electron operator.\cite{Fiete:prl08,Fiete:prb10})  Careful studies of the exchange anisotropy have shown that the anisotropy is {\em irrelevant} only for $S=1/2$ and $S=(k-1)/2$.\cite{Affleck:prb92}  Thus, for $S=1$ the multi-channel non-Fermi liquid behavior is only to be expected for $k=3$, or for the $\nu=13/5$ plateau, if it appears in experiment and is described by a RR state. 
However, the $S=1$ non-Fermi liquid case suffers from the usual instabilities if one tunes away from the special point of equal $t,\epsilon,V$ couplings:  It will generically flow to a Fermi liquid fixed point.\cite{Affleck:prb92}

If we generalize the remote double-dot case to multiple remote dots (that is, those that do not directly interact), it is clear that for $N$ identical dots, one will have a low-energy fixed point described by a $S=N/2$ $k$-channel Kondo model, which, because of the intrinsic exchange anisotropy\cite{Affleck:prb92} will exhibit Fermi liquid physics unless $N/2=(k-1)/2$ or $k=N+1$ and all the couplings are fine-tuned to identical values of each dot.  We thus conclude that the ``protected" $k$-channel Kondo non-Fermi liquid physics of the single-dot (composite dot) limit appears to be rather special to the single-dot limit with a 2-fold occupancy degeneracy.\cite{Fiete:prl08,Fiete:prb10}  However, there are other possibilities when indirect dot-dot interactions are included.

\subsubsection{Effective inter-dot interactions}

Our analysis of the the low-energy fixed points of the double and mult-dot systems did not explicitly include indirect inter-dot interactions that would be generated perturbatively, and these can be important for the low-energy fixed point.\cite{Gan:prl95,Gan:prb95,Jones:prl87,Jones:prl88,Affleck:prl92,Affleck:prb95}   In the 2-impurity Kondo model in a Fermi liquid, these perturbatively generated interactions often go under the name of RKKY-interactions.\cite{Ruderman:rmp54}  Here we will derive their analog in the double-dot case on the quantum Hall edge and discuss their effect on the physics.  

At energy scales above those close to the fixed point, but below the quantum Hall bulk gap, the relevant form of the Hamiltonian is 
\begin{eqnarray}
\label{eq:2_remote}
H=H_{\rm edge}+\sum_{i=1}^2 t_i\left(\Psi^{\dagger}_{e}(x_i) d_i+d_i^{\dagger} \Psi_{e}(x_i)\right)\nonumber \\
+\sum_{i=1}^2\epsilon_i n_i+\sum_{i=1}^2  V_i n_i\partial_x\phi(x_i), 
\end{eqnarray}
where now that we are working at higher energies, we take $x_1 \neq x_2$. Since the dots are assumed remote, we have $t_{ij}=V_{ij}=0$.

The Hamiltonian \eqref{eq:2_remote} can be cast into the form of a 2-impurity Kondo model upon application of the unitary transformation $U=e^{i\alpha\sum_i^2 S_i^z\phi(x_i)}$ with $\alpha=\frac{k+2}{k}-\frac{\sqrt{2(k+2)}}{k}$ for the general RR state with electron operator given by \eqref{eq:Psi_RR}:
\begin{eqnarray}
\label{eq:2_U_Kondo}
UHU^\dagger=H_{\rm edge}+\sum_{i=1}^2\lambda_{i\perp}\left(J^+(x_i)S_i^-+J^-(x_i)S_i^+\right) \nonumber \\
+\sum_{i=1}^2 h_iS_i^z+\sum_{i=1}^2\lambda_{iz}J^z(x_i)S_i^z,\;\;\;
\end{eqnarray}
where $\lambda_{i\perp}=t_i$, $\beta \lambda_{iz}=V_i-{v_c}\alpha$, $\beta=\sqrt{2(k+2)}/k$, $h_i=\epsilon_i$,
and the $SU(2)_k$ currents are given by \eqref{eq:SU2_currents} as before. The components of the spin operators $S_i$ are  $S_i^+ =\eta_i d_i^\dagger, S_i^-=d_i\eta_i$,
and $S_i^z=n_i -1/2$, where $\eta_i$ are dot-dependent Klein factors that ensure the proper commutation relations are achieved.\cite{Klein_factors} We remark that the transformation between \eqref{eq:2_remote} and \eqref{eq:2_U_Kondo} is {\em exact}, and the physics should be equivalently captured by either form of the Hamiltonian. The most important difference between the two forms is that the coupling constants $\lambda_{i\perp},\lambda_{iz}$ have the same scaling dimensions ({\em i.e.} marginal) in \eqref{eq:2_U_Kondo} while in \eqref{eq:2_remote} the tunneling terms are {\em naively} irrelevant.\cite{Fiete:prl08,Fiete:prb10}

The Hamiltonian \eqref{eq:2_U_Kondo} shows that at ``high" energies two spatially separated quantum dots behave as a 2-impurity exchange anisotropic but channel isotropic $k$-channel Kondo model in an gapless electron liquid described by $H_{\rm edge}$.  To the best of our knowledge, this situation has not arisen before in the literature. The physics of 2-impurity $k$-channel models in the case of a Fermi liquid has not been studied, aside from the case of $k=2$ which has received only limited attention.\cite{Ingersent:prl92,Georges:prl95}  We thus believe there is the potential for novel features to arise in this class of impurity models.

As a point of comparison, it is useful to briefly remind the reader of the formulation of the 2-impurity problem in the Fermi liquid context.  The following Hamiltonian plays a prominent role,\cite{Gan:prl95,Gan:prb95,Jones:prl87,Jones:prl88,Affleck:prl92,Affleck:prb95}
\begin{equation}
\label{eq:2_imp_FL}
H_{\rm 2-imp\;FL}=\lambda\vec{J}(x_1) \cdot \vec{S_1}+ \lambda \vec{J}(x_2) \cdot \vec{S_2}+K \vec{S_1} \cdot \vec{S_2},
\end{equation}
where the spin-spin interactions with coupling constant $K$ are typically generated by RKKY-type effects.\cite{Ruderman:rmp54}  The spin-spin interactions can be computed perturbatively in $\lambda$ in \eqref{eq:2_imp_FL} so that $K \propto \lambda^2$ at lowest order.  However, for purposes of studying possible phase transitions, it is useful to think of $K$ as in independent parameter that can be tuned (by changing the impurity separation, for example), as it will in any case be renormalized at low energies.  Ultimately, the physics is determined by a complex interplay between both single-impurity-like Kondo effects and the impurity-impurity interactions. 

The main result of the studies of the single-channel model \eqref{eq:2_imp_FL} is that it possess a Fermi-liquid ground state except for the case where particle-hole symmetry is present {\em and} $K$ takes on a special value of order $T_K$, where $T_K$ is the single-impurity Kondo temperature.\cite{Gan:prl95,Gan:prb95,Jones:prl87,Jones:prl88,Affleck:prl92,Affleck:prb95}  One can argue for a phase transition in the following way.  When $K\to -\infty$ the two spins are ferromagnetically coupled and effectively behave as $S=1$, with two channels (one from each spin-1/2) coupled to the conduction electrons.  Since the number of channels is equal to 2 x $1/2$, the ground state is Fermi liquid-like and there is a unitary limit scattering phase shift of $\pi/2$ at low energies.  On the other hand, when $K\to \infty$, the two spins are locked into a singlet and thus ``screen" each other resulting in a phase shift of zero. (They become inert from the point-of-view of the conduction electrons.)  Because the phase shift cannot evolve smoothly between $0$ and $\pi/2$ there must be a phase transition in between for some intermediate value of $K$; it turns out that right at the critical point the ground-state has non-Fermi liquid properties.\cite{Gan:prl95,Gan:prb95,Jones:prl87,Jones:prl88,Affleck:prl92,Affleck:prb95}

The 2-impurity, 2-channel Kondo model in a Fermi liquid has been studied in Refs.[\onlinecite{Ingersent:prl92},\onlinecite{Georges:prl95}].  The main result is that the single-impurity non-Fermi liquid behavior of the 2-channel model is destabilized by the RKKY interactions.  In addition to a number of Fermi-liquid phases,  a line of non-universal, non-Fermi liquid ground states is found for antiferromagnetic $K$.  The manifold of non-Fermi liquid ground states with properties that vary with the bare parameters results from the RKKY interaction being an exactly marginal perturbation in this case.  Thus, non-Fermi liquid physics is more stable in the 2-impurity, 2-channel Kondo model than in the 2-impurity, single-channel Kondo model (which has only one {\em unstable} point in parameter space of non-Fermi liquid behavior). 

We now return to the subject of the RKKY interactions for the Hamiltonian \eqref{eq:2_U_Kondo}.  To the best of our knowledge, RKKY interactions have not been investigated in an interacting chiral system such as the edge of a fractional quantum Hall state.  Even in a Luttinger liquid, they appear to have received little attention.\cite{Egger:prb96}  The most salient difference is that the Fermi wave vector $k_F$ does explicitly appear in the edge electron operators, \eqref{eq:Psi_Laugh}, \eqref{eq:Psi_MR}, or \eqref{eq:Psi_RR} [or the closely related $SU(2)_k$ current operators, \eqref{eq:SU2_currents}].  This has the immediate consequence that the RKKY interactions for  \eqref{eq:2_U_Kondo} will not be oscillatory.  

The sign of the interaction can be determined by energetics: Considering first the $S^z$ interactions, if an electron is on dot 1, but not on dot 2, then a second order process of an electron hopping to dot 2 and back will be possible while it would not be if dot 2 were occupied.  Thus, the effective interaction
\begin{equation}
\label{eq:RKKY_z}
H_{\rm RKKY, z-comp}=K^z(|x_2-x_1|)S_1^zS_2^z,
\end{equation}
is antiferromagnetic, $K^z>0$, for all distances.  Because of translational symmetry along the edge, the interaction only depends on the distance between the dots.  In the language of dot occupations, the Hamiltonian \eqref{eq:RKKY_z} is a density-density interaction between the dots.

On the other hand, if one looks at the perpendicular components, one finds
\begin{equation}
\label{eq:RKKY_perp}
H_{\rm RKKY, \perp-comp}=K^\perp(|x_2-x_1|)\left(S_1^+S_2^-+{\rm H.c.}\right),
\end{equation}
which corresponds to a {\em tunneling} term between dots. Since $K^\perp(x)=\lambda_{1,\perp}\lambda_{2,\perp}\langle J^+(x)J^-(0)\rangle$ and $K^z(x)=\lambda_{1,z}\lambda_{2,z}\langle J^z(x)J^z(0)\rangle$ are proportional by the $SU(2)$ symmetry of the edge currents, the perpendicular components are also antiferromagnetic.  Moreover, as the scaling dimensions of the current operators are independent of $k$, we find $K^{z,\perp}(x)\propto 1/x^2$ for all quantum Hall edges in the RR class.  Thus, there is a universal decay of the RKKY interactions in the double dot scenario we consider.

If we had instead computed the effective dot tunneling terms $\tilde K^\perp\propto t_1t_2$ from \eqref{eq:2_remote} we would have found $\tilde K^\perp(x) \propto 1/x^3$ so that we would have concluded that these are sub-leading terms compared to $K^z$.  However, as discussed in Refs[\onlinecite{Fiete:prl08},\onlinecite{Fiete:prb10}] the presence of the dot-edge interaction, $V$, changes the scaling dimension of the $t_i$ to match $V$ at low-energies.  Thus, for studies of the possible low-energy fixed points one should use the form \eqref{eq:2_U_Kondo} which already includes the proper scaling dimension of the terms proportional to $t_i$.\cite{Fiete:prl08,Fiete:prb10}

Finally, we note that if one goes to fourth order perturbation theory in $\lambda_\perp$, density-density interactions similar to \eqref{eq:RKKY_z} will result.  For the effective $S=1/2$ model we are considering here, no other interactions will be generated at higher orders in perturbation theory.  We thus see that the Hamiltonian of the dots with RKKY interactions included reduces to a form similar to \eqref{eq:H_QD_2}, which we had considered earlier.

Since the effective spin-spin interactions \eqref{eq:RKKY_z} and \eqref{eq:RKKY_perp} are antiferromagnetic, analogy to the $k=2$ Fermi liquid case\cite{Ingersent:prl92,Georges:prl95} suggest that there is a possibility for new non-Fermi liquid fixed points to arise.  We leave a detailed study of this intriguing possibility to future work.

\section{Conclusions}
\label{sec:conclusions}

In this work we have studied electron tunneling between multiple quantum dots and the edge of a non-Abelian fractional quantum Hall state.  We have investigated how much of the exact mapping\cite{Fiete:prl08,Fiete:prb10} of a single dot next to a Read-Rezayi state at filling fraction $\nu=2+k/(k+2)$ to a $k$-channel Kondo model carries over to various multiple dot cases. We did not study particle-hole conjugates of the Read-Rezayi states or other candidate states in the literature (non-Abelian as well as Abelian) as they generally result in Fermi liquid quantum impurity physics.\cite{Fiete:prl08,Fiete:prb10}  We stress that the multiple impurity $k$-channel Kondo model has not been addressed before in the literature, aside from the case of $k=2$ in a Fermi liquid.\cite{Ingersent:prl92,Georges:prl95}  The typically fragile non-Fermi liquid physics of the single dot (spin) case is ``topologically protected" in our scenario and we believe this opens the way for potentially a new class of fixed points, perhaps analogous to what is found in $k=2$ case,\cite{Ingersent:prl92,Georges:prl95} or something altogether different.  
Motivating further studies on the models we consider here and related models is one of the primary goals of this work.

In this paper we established that in many situations, such as dots in close proximity, the multiple dots may act as one effective dot.  In this case, if the energy levels are tuned to a {\em double} degeneracy point, then the charge fluctuations map onto the single-impurity $S=1/2$ $k$-channel Kondo model as before.  However, we also found that higher degeneracies are possible, and these correspond to higher-spin multi-channel Kondo models.  We showed that in general the non-Fermi liquid physics of these higher spin multi-channel Kondo models is {\em not} topologically protected, and noted that they will instead generically exhibit Fermi liquid properties.  In this sense, the ``single dot" limit previously studied captures a large class of the interesting non-Fermi liquid impurity physics.\cite{Fiete:prl08,Fiete:prb10}

We also studied the case of impurities widely separated along the edge (so that they do not directly interact with each other).  Specializing to the case of two widely separated impurities, we computed the effective RKKY-type interactions between the charges on the dots.  These terms are known to potentially compete with the single-impurity Kondo physics and drive phase transitions, including a non-Fermi liquid fixed point in the case of two single-channel Kondo impurities in a Fermi liquid,\cite{Gan:prl95,Gan:prb95,Jones:prl87,Jones:prl88,Affleck:prl92,Affleck:prb95} and a line of non-universal non-Fermi liquid behavior in the case of two 2-channel Kondo impurities in a Fermi liquid.\cite{Ingersent:prl92,Georges:prl95} We speculate that even more non-Fermi liquid states are to be found in the scenario we discussed, but leave that question to future work.  In particular, if numerical renormalization group methods\cite{Kim:prb97,Bulla:rmp08} can be adapted to handle the strongly correlated gapless degrees of freedom present on the edge of a non-Abelian fractional quantum Hall state, then significant progress could be made in this direction. 

\begin{figure}[h]
\includegraphics[width=0.95\linewidth,clip=]{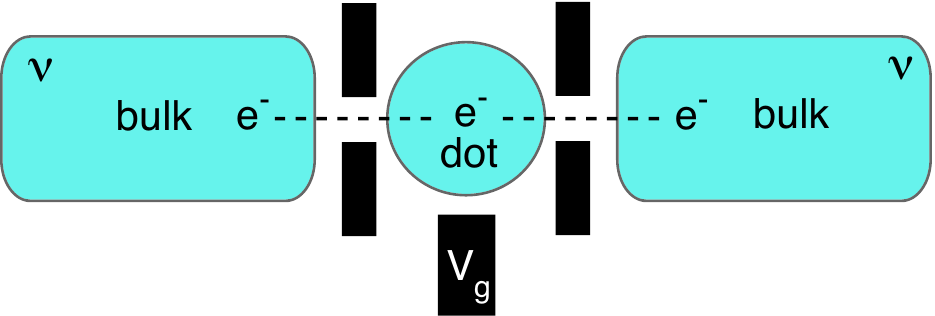}
 \caption{\label{fig:transport} (color online) Schematic of a model for electron tunneling through a quantum dot.  Gates are shown in black.  The gates can be used to form a point contact to pinch the dot off from the rest of the quantum Hall bulk.  The gate labelled $V_g$ can be used to adjust the number of electron on the dot.  When electron tunneling is allowed between the dot and the two reservoirs, the charge fluctuations near a degeneracy point have physics that does not immediately map onto a Kondo impurity model and a novel quantum impurity fixed point is possible.}
 \end{figure}

The multi-dot scenarios we studied in this work were all edge-coupled to the fractional quantum Hall system, as shown in Fig.\ref{fig:schematic}.  Another possibility is to have a single dot or multiple dots connected in series to two fractional quantum Hall states, as shown in Fig.\ref{fig:transport}.  Remarkably, even the case of a single quantum dot shown in Fig.\ref{fig:transport} does not appear to have a simple mapping to any Kondo-like quantum impurity problem when the quantum Hall states are {\em fractional} (including a simple Lauglin state).\cite{Fiete:prb10}  We hold this system out as yet another where novel quantum impurity fixed points may arise.

In summary, our main goal in this work was to understand how much of the non-Fermi liquid impurity physics of a single quantum dot near its degeneracy point  carries over to the case of multiple dots.  We outlined some conditions under which we expected the non-Fermi liquid physics to be: (i) ``topologically protected" as before, and (ii) present for a fine-tuning of parameters.  However, there are many details left unanswered by our study and possibly new fixed points to be discovered.  It is our hope that other researchers will pick up the thread and apply the full power of conformal field theory, non-perturbative renormalization group methods, and other techniques to study this problem further.  We believe there is potential for fundamental discovery in quantum impurity physics in the scenarios we have discussed here.

\acknowledgements
GAF is grateful to Waheb Bishara and Chetan Nayak for collaborations on earlier related works, and to Victor Galitski for discussions.  This work was supported in part by ARO grant W911NF-09-1-0527 and NSF grant DMR-0955778.  SAS gratefully acknowledges support from an Intel Foundation Research Fellowship and an NSF Graduate Research Fellowship.

%

\end{document}